\documentclass[11pt]{amsart}

\usepackage{amssymb,amsthm,amsmath}
\usepackage{fullpage}
\usepackage[all]{xy} 

\newtheorem{thm}{Theorem}[section]
\newtheorem{prop}[thm]{Proposition}
\newtheorem{lemma}[thm]{Lemma}
\newtheorem{cor}[thm]{Corollary}

\theoremstyle{definition}
\newtheorem{defn}[thm]{Definition}
\newtheorem{rem}[thm]{Remark}

\newtheorem{prot}[thm]{Protocol}
\newtheorem{conj}[thm]{Conjecture}

\numberwithin{equation}{section}

\newcommand{\F}{\mathbb{F}}

\newcommand{\Z}{\mathbb{Z}}
\newcommand{\G}{\mathbb{G}}
\newcommand{\A}{\mathcal{A}}
\newcommand{\B}{\mathcal{B}}
\newcommand{\C}{\mathcal{C}}
\renewcommand{\P}{\mathcal{P}}
\newcommand{\V}{\mathcal{V}}
\newcommand{\GG}{\mathcal{G}}
\newcommand{\bit}[1]{\{0,1\}^{#1}}

\DeclareMathOperator{\Adv}{Adv}

\newcommand{\abs}[1]{\left| #1 \right|}

\begin{document}

\title{Pairing-based identification schemes}
\author{David Freeman}
\address{University of California, Berkeley}
\email{dfreeman@math.berkeley.edu}
\begin{abstract}
We propose four different identification schemes that make use of bilinear pairings, and prove their security under certain computational assumptions.  Each of the schemes is more efficient and/or more secure than any known pairing-based identification scheme.
\end{abstract}
\maketitle

\section{Introduction}
\label{s:intro}

An {\it identification scheme} is a protocol whereby Peggy the Prover proves to Victor the Verifier that she is indeed who she says she is.  In practice, Peggy's identity is encoded in a private key $a$ and a public key $y$.  The protocol takes the form of Peggy proving to Victor that she has knowledge of the private key $a$.  For example, the private key might be $a$ and the public key $y = x^a \pmod{p}$, where $a$ and $x$ are integers and $p$ is a prime number, and Peggy proves her identity by demonstrating that she knows the discrete logarithm of $y$ to the base $x$.  Now, Peggy could simply tell Victor $a$, and Victor could verify that $a$ is the correct private key, but then Victor could impersonate Peggy to a third party.  A viable identification scheme must prevent this from happening; we require that Victor can't impersonate Peggy even if she proves her identity to him polynomially many times.  Because of this property, an identification scheme is also called a {\it zero-knowledge proof of identity}.

Feige, Fiat, and Shamir \cite{ffs} introduced the first identification scheme in 1988, based on the difficulty of inverting RSA.  Soon thereafter, Guillou and Quisquater \cite{gq} and Schnorr \cite{sch} introduced their own identification schemes, based on RSA and discrete logarithms respectively.  These two schemes are still amongst the most efficient and well-studied identification schemes, though their security has never been reduced to a standard computational problem such as factoring or discrete logarithms.

Identification schemes are closely related to signature schemes.  For example, one way for Peggy to prove her identity to Victor is for him to ask her to digitally sign a message of his choice; if the signature is hard to forge, then a valid signature will constitute an acceptable proof of identity.  On the other hand, many of the standard identification schemes can be converted to a signature scheme by replacing Victor with a one-way hash function.

Recent years have brought a host of signature schemes that make use of bilinear pairings.  The first of these was the short signature scheme of Boneh, Lynn, and Shacham in 2001 \cite{bls}.  This was quickly followed by a spate of pairing-based schemes designed for various applications: group signatures, ring signatures, aggregate signatures, multisignatures, threshold signatures, and more.  Given this plethora of pairing protocols and the close relationship between identification schemes and signatures, it is natural to ask whether there might be a pairing-based identification scheme that has some advantage over the GQ or Schnorr schemes.  The first step in this direction was taken by Kim and Kim in 2002 \cite{kk}.  Their scheme was later shown to be flawed; others have since proposed pairing-based identification schemes \cite{hls}, \cite{scl}, \cite{yww}, but none has given a convincing proof of security with a tight reduction.

In this paper, we present four new identification schemes based on pairings, and prove their security given certain computational assumptions.  We begin in Section \ref{s:idscheme} by giving a formal definition of security for identification schemes, reviewing some standard computational assumptions, and describing the bilinear pairings useful for cryptography.  In Section \ref{s:bls}, we describe a basic scheme based on the Boneh-Lynn-Shacham signatures and prove its security in the random oracle model under the Computational Diffie-Hellman assumption.  Since the random oracle model is somewhat unsatisfactory for proving security of identification schemes, in Section \ref{s:cdh} we modify the scheme so that it does not require the use of hash functions.  To prove security of this new scheme we introduce a new assumption, called the ``one-more-Computational Diffie-Hellman'' assumption, which is related to several existing assumptions in the literature.

In Section \ref{s:sdh} we take another tack, adapting a signature scheme that does not make use of random oracles for its proof of security.  The proof of security of this scheme relies of the ``Strong Diffie-Hellman assumption,'' an analogue of the ``Strong RSA assumption'' used to prove security of RSA signatures.  Finally, in Section \ref{s:owf} we introduce a scheme whose proof of security relies on the assumption that the pairing used is a one-way function.  We show that this assumption is weaker than any other made in this paper, and thus this scheme is the most secure of our new schemes.

Having presented our four new schemes and proved their security, in Section \ref{s:other} we describe two other pairing-based identification schemes in the literature, and in Section \ref{s:comparison} we examine the bandwidth and computational requirements of all six schemes.  We conclude that each of our four protocols is the preferred identification scheme in some context, for either efficiency or security reasons.

\subsection{Acknowledgments}

Research for this paper was conducted during a summer internship at HP Labs, Palo Alto.  I thank Vinay Deolalikar for suggesting this problem and for providing advice and support along the way.  I also thank Gadiel Seroussi for bringing me to HP and for supporting my research.

\section{Preliminaries}
\label{s:idscheme}

\subsection{Identification schemes}

Formally, an identification scheme consists of a key-generation algorithm $\mathcal G$ that creates a valid set of keys $a$ (Peggy's private key) and $p_a$ (Peggy's public key), and an interactive protocol $(\P,\V)$ that takes as input the public and private keys, and outputs 1 (accept) or 0 (reject).  We require that if both users follow the protocol and use a valid public/private key pair, the protocol always outputs 1 (accepts).  We also require that any cheating prover $\A$ that does not know Peggy's private key cannot interact with an honest verifier $\V$ and give output 1; this is a ``passive attack.''  Furthermore, we require that a cheating verifier $\B$ cannot interact with Peggy, pass what he learns on to the cheating prover $\A$, and have $\A$ interact with an honest verifier $\V$ and output 1; this is an ``active attack.''  We note that a passive attack is a special case of an active attack, in which $\B$ outputs nothing.  This leads us to the following definition:

\begin{defn}[{cf.\ \cite[Definition 4.7.8]{gold}}]
\label{d:idscheme}
A $(t,q,\epsilon)$-{\it identification scheme} is a triple $(\GG,\P,\V)$, where $\GG$ is a probabilistic polynomial-time algorithm and $(\P,\V)$ is a pair of probabilistic interactive machines running in time at most $t$, satisfying the following conditions:
\begin{itemize}
\item {\it Viability:}  For any $\alpha \in \bit{n}$, let $\GG(\alpha) = (a_\alpha,p_\alpha)$.  Then
$$\Pr\left[\langle \P(a_\alpha,p_\alpha), \V(p_\alpha) \rangle = 1\right] = 1.$$
\item {\it Security:} For any $\alpha \in \bit{n}$, let $\GG(\alpha) = (a_\alpha,p_\alpha)$.  For any probabilistic interactive machine $\B$ running in time at most $t$, let $T_\alpha$ be a random variable describing the output of $\B(p_\alpha)$ after interacting with $\P(a_\alpha,p_\alpha)$ $q$ times.  Then for any probabilistic interactive machine $\A$ running in time at most $t$, 
$$\Pr\left[\langle \A(p_\alpha,T_\alpha), \V(p_\alpha) \rangle = 1\right] < \epsilon.$$
\end{itemize}
\end{defn}

Note that the security condition implies that a third party, Malice, cannot impersonate Peggy to Victor, {\it provided that Malice cannot interact concurrently with Peggy and Victor}.  Indeed, if Malice can interact concurrently with both, she may impersonate Peggy by referring Victor's queries to Peggy and relaying the response back to Victor.

\subsection{Computational assumptions}

All of public-key cryptography relies on certain computational assumptions for its security; e.g.\ that factoring is difficult.  The assumptions relevant to our identification schemes are of the {\it Diffie-Hellman} type, named after the two creators of public-key cryptography.  The original Diffie-Hellman problem is known as the Computational Diffie-Hellman (CDH) problem.

\begin{defn}
\label{d:cdh}
Let $\G$ be a cyclic group of order $n$, let $g \in \G$, and let $a,b \in \Z_n$.  The {\it Computational Diffie-Hellman problem} in $\G$ is as follows: Given $\{g,g^a,g^b\}$, compute $g^{ab}$.

The {\it $(t,\epsilon)$-Computational Diffie-Hellman assumption} holds in $\G$ if there is no algorithm $\A : \G^3 \to \G$ running in time at most $t$ such that
$$\Pr \left[ \A(g,g^a,g^b) = g^{ab} \right] \geq \epsilon , $$
where the probability is taken over all possible choices of $(g,a,b)$.
\end{defn}

It is possible that given a triple $(g,g^a,g^b)$, it is hard to compute $g^{ab}$ but easy to compute some partial information about $g^{ab}$, such as its least significant bit.  To ensure that no such partial information can be gained, we must make an even stronger assumption, known as the Decision Diffie-Hellman (DDH) assumption.  

\begin{defn}
\label{d:ddh}
Let $\G$ be a cyclic group of order $n$, let $g \in \G$, and let $a,b,c \in \Z_n$.  The {\it Decision Diffie-Hellman problem} in $\G$ is as follows: Given $\{g,g^a,g^b,g^c\}$, determine whether $g^{ab} = g^c$.

The {\it $(t,\epsilon)$-Decision Diffie-Hellman assumption} holds in $\G$ if there is no algorithm $\A : \G^4 \to \bit{}$ running in time at most $t$ such that
$$\abs{\Pr \left[ \A(g,g^a,g^b,g^{ab}) = 1 \right] - \Pr \left[ \A(g,g^a,g^b,g^c) = 1 \right] } \geq \epsilon, $$
where the probabilities are taken over all possible choices of $(g,a,b,c)$.
\end{defn}

\subsection{Bilinear maps and pairings}

Joux and Nguyen \cite{jn} showed that an efficiently computable bilinear map on $\G$ gives an algorithm for solving the Decision Diffie-Hellman problem on $\G$.  Boneh, Lynn, and Shacham \cite{bls} make use of this property in their signature algorithm by using the pairing to verify that the signature creates a valid Diffie-Hellman tuple.  Our identification schemes will use pairings in their verification procedures in a similar manner.

The following definition gives the conditions necessary for a bilinear map to be useful for cryptographic purposes. To simplify our exposition, we will consider only the case where both arguments of the pairing are in the same group; for the more general case, see \cite{bls}.

\begin{defn}
\label{d:pairing}
Let $\G_1$ and $\G_2$ be cyclic groups of prime order $p$.  A map $e \colon \G_1 \times \G_1 \to \G_2$ is a {\it cryptographic pairing} if the following conditions hold:
\begin{itemize} 
\item {\it Bilinearity:} for all $x,y \in \G_1$ and $a,b \in \Z$, $e(x^a,y^b) = e(x,y)^{ab}$.
\item {\it Non-degeneracy:} if $g$ is a generator of $\G_1$, then $e(g,g)$ is a generator of $\G_2$.
\end{itemize}
\end{defn}

\begin{rem}
\label{r:gdh}
A cryptographic pairing $e$ can be used to solve the DDH problem on $\G_1$ as follows: given $\{g,g^a,g^b,g^c\}$, where $g$ is a generator of $\G_1$ and $a,b,c$ are integers, compute $h_1 = e(g,g^c)$ and $h_2 = e(g^a,g^b)$.  Then $h_1 = h_2$ in $\G_2$ if and only if  $c = ab \pmod{p}$.  If the CDH problem in $\G_1$ is hard and the DDH problem is easy (e.g.\ if there is a cryptographic pairing on $\G_1$), $\G_1$ is known as a {\it Gap Diffie-Hellman group}.  The {\it Gap Diffie-Hellman problem} is to solve the CDH problem given an oracle for the DDH problem.
\end{rem}

The only known examples of cryptographic pairings are derived from the Weil and Tate pairings on elliptic curves over finite fields.  The study of these groups is deep and beautiful and is of great interest to current researchers.  However, in describing our protocols we will not take into account the structure of the groups involved in the pairing; rather, we will make certain computational assumptions about the group and use the pairing as a ``black box.''  For further information on elliptic curves, see \cite{bss} or \cite{bss2}.

\section{Identification scheme based on BLS signatures}
\label{s:bls}

A particularly simple method of building identification schemes is to use a digital signature algorithm.  Victor the Verifier sends a random message to Peggy the Prover, Peggy signs the message with her secret key, and Victor verifies that the signature is correct.  If the signature scheme is secure against forgery, the cheating prover has a negligible chance of creating a valid signature on a random message given him by an honest verifier, no matter how many signatures he has obtained from the honest prover.

Boneh, Lynn, and Shacham \cite{bls} were the first to devise a digital signature scheme based on pairings.  The algorithm provides for signatures of half the length of a DSS signature with an equivalent level of security, and as such it makes for a particularly efficient identification scheme in terms of bandwidth.  A full description of the BLS signature scheme, along with a definition of security for signature schemes and the security theorem for the BLS scheme, can be found in Appendix \ref{a:signatures}.

We now show how the BLS signature scheme can be adapted nearly verbatim to serve an an identification scheme.  We describe the scheme as an interactive protocol between Peggy the prover and Victor the verifier.

\begin{prot}
\label{p:blsid}
Let $\G_1$, $\G_2$ be cyclic groups of prime order $p$, and let $e \colon \G_1 \times \G_1 \to \G_2$ be a cryptographic pairing.  Let $g$ be a generator of $\G_1$.  Let $H : \bit{*} \to \G_1$ be a full-domain hash function.
\begin{description}
\item[Key generation] Pick random $x \leftarrow \Z_p$, and compute $v \leftarrow g^x$.  The public key is $v$, and Peggy's secret key is $x$.  Let $n$ be a positive integer.
\item[Interactive protocol] \ 
\begin{enumerate}
    \item Victor sends Peggy a random $M \in \bit{n}$.
    \item Peggy computes $h = H(M)$ and sends Victor $\sigma = h^x$.
    \item Victor computes $e(g,\sigma)$ and $e(v,h)$.  If the two are equal he outputs $1$ (accept); else he outputs $0$ (reject).
\end{enumerate}
\end{description}
\end{prot}

Since our signature makes use of a hash function and the proof of security is in the random oracle model, we must add another parameter to our description of security of identification schemes.  We say that a scheme using a hash function is a $(t,q,r,\epsilon)$-identification scheme if the conditions of Definition \ref{d:idscheme} hold, with the additional requirement that $(\A,\B)$ make no more than $r$ queries to the hash function.

\begin{thm}
\label{t:blsid}
Suppose the $(t',\epsilon')$ Computational Diffie-Hellman assumption holds in $\G_1$.  Then Protocol \ref{p:blsid} defines a $(t,q_S,q_H,\epsilon)$-identification scheme for all $t$ and $\epsilon$ satisfying
$$\begin{array}{ccc}
\epsilon \geq \displaystyle{ \frac{2^n e(q_S+1)}{2^n - q} \cdot \epsilon' } & \mbox{and} & t \leq t' - c(q_H + 2q_S),
\end{array}$$
where $c$ is a constant that depends on $\G_1$, and $e$ is the base of the natural logarithm.
\end{thm}

\begin{proof}[{\bf Proof (sketch)}]
If Peggy and Victor follow the protocol, then Protocol \ref{p:blsid} satisfies the viability condition of Definition \ref{d:idscheme}, since
$$e(g,\sigma) = e(g,h^x) = e(g,h)^x = e(g^x,h) = e(v,h)$$
by bilinearity of $e$.  The security follows from the security of the BLS scheme: a successful cheating prover $\A$ will send an element $\sigma$ in step (2) that is accepted by the honest verifier.  This $\sigma$ is, with high probability, a valid BLS signature for a previously unseen message $M$.  The security of the BLS scheme against existential forgery under chosen-message attack thus implies the security of Protocol \ref{p:blsid}.  The exact bounds for the running time and success probability follow from the proof of security of the BLS scheme (Theorem \ref{t:bls}).  For details, see Appendix \ref{a:blsproof}.
\end{proof}

\section{Identification schemes based on the one-more-CDH assumption}
\label{s:cdh}

Protocol \ref{p:blsid}, an identification scheme derived directly from the BLS signature scheme, is unsatisfactory in several ways.  While the communication overhead is minimal (one element of $\G_1$ and one random string which needs only to be large enough to avoid hash collisions), the prover and verifier must both compute the hash of the parameter $M$, which adds computational time.  In addition, the proof of security is in the random oracle model, which requires us to introduce another security parameter and to assume that the hash function $H$ acts as a random function.  Recent attacks on SHA-1 and other hash functions have called into question the credibility of such an assumption, so we would ideally like our identification schemes to be hash-free.

Our first attempt at constructing a pairing-based identification scheme that does not use hash functions is simply to recreate the scheme based on BLS signatures, but do away with the hash function.

\begin{prot}
\label{p:cdhid}
Let $\G_1$, $\G_2$ be cyclic groups of prime order $p$, and let $e \colon \G_1 \times \G_1 \to \G_2$ be a cryptographic pairing.  Let $g$ be a generator of $\G_1$.
\begin{description}
\item[Key generation] Pick random $x \leftarrow \Z_p$, and compute $v \leftarrow g^x$.  The public key is $v$, and Peggy's secret key is $x$.
\item[Interactive protocol] \ 
\begin{enumerate}
    \item Victor sends Peggy a random challenge $h \in \G_1$.
    \item Peggy computes sends Victor $\sigma = h^x$.
    \item Victor computes $e(g,\sigma)$ and $e(v,h)$.  If the two are equal he outputs $1$ (accept); else he outputs $0$ (reject).
\end{enumerate}
\end{description}
\end{prot}

We can think of Protocol \ref{p:cdhid} as Protocol \ref{p:blsid} where instead of sending a random message $M$ in step (1), Victor sends the hash $h$ of the message $M$; if the hash is random, then $h$ is just a random element of $\G_1$.  With this modification, the reduction of the scheme to the Computational Diffie-Hellman assumption in $\G_1$ breaks down, as that reduction requires that Peggy can't compute $M$ from $h$.  The security of this scheme thus requires a different assumption.

To determine what kind of security assumption we need to make, we examine the behavior of an attacker.  The cheating verifier $\A$ interacts with the honest prover $\P$ by sending $q$ queries of her choice $h_1,\ldots,h_q$ and receiving the `signature' of each message, $h_1^x,\ldots,h_q^x$.  The cheating prover $\B$ must then take a random query $h$ and return $h^x$.  (Note that by the bilinearity of the pairing $e$, $h^x$ is the only element that $\B$ can send in step (2) that will cause an honest verifier to accept.)  If $q = 0$, then this is the Computational Diffie-Hellman problem: compute $h^x$ from $\{g,g^x,h\}$.  If $q > 0$, we are asking for the solution to a CDH problem given the solution to $q$ related CDH problems.  We formalize this notion in the following definition.

\begin{defn}
\label{d:omcdh}
Let $\G$ be a finite cyclic group.  Let $A$ be a randomized algorithm that takes input $g,g^a \in \G$ and has access to two oracles.  The first is a CDH oracle $CDH_{g,g^a}(\cdot)$, which on input $h \in \G$ returns $h^a \in \G$.  The second is a challenge oracle $C()$ that, when invoked, returns a random challenge point $r \in \G$.  Furthermore, we require that $\A$ cannot invoke its CDH oracle after it has invoked the challenge oracle.
We say that algorithm $\A$ has advantage $\epsilon$ in solving the {\it one-more-CDH problem} in $\G$ if
$$\Pr \left[ \A(g,g^a, r \leftarrow C() ) = r^a \right] \geq \epsilon,$$
where the probability is taken over the choices $g$ and $g^a$ input to $\A$ and the $r$ output from $C()$.

We say the {\it $(t,q,\epsilon)$-one-more-CDH assumption} holds in $\G$ if there is no algorithm $\A$ that runs in time at most $t$, makes at most $q$ queries to its CDH oracle, and has advantage at least $\epsilon$ in solving the one-more-CDH problem in $\G$.
\end{defn}

Definition \ref{d:omcdh}, while it has not appeared previously in the literature, is closely related to the ``one-more-RSA-inversion'' and ``one-more-discrete-logarithm'' problems defined by Bellare, et al.\ \cite{bnps}.  Bellare and Palacio \cite{bp} use these assumptions to prove the security of the well-known Guillou-Quisquater and Schnorr identification schemes, so it seems eminently reasonable that we should have to use a similar assumption in proving the security of our scheme.

We now prove the security of Protocol \ref{p:cdhid} based on the one-more-CDH assumption.

\begin{thm}
\label{t:cdhid}
Suppose the $(t,q,\epsilon)$-one-more-CDH assumption holds in $\G$.  Then Protocol \ref{p:cdhid} is a $(t-O(1),q,\epsilon)$-identification scheme.
\end{thm}

\begin{proof}
Let $(g,g^x)$ be the public parameters for Protocol \ref{p:cdhid}.  Suppose $(\A,\B)$ is an attack that $(t,q,\epsilon)$-breaks Protocol \ref{p:cdhid} in the sense of Definition \ref{d:idscheme}.  Define an algorithm $\C$ that attempts to solve the one-more-CDH problem in $\G_1$, as follows:
\begin{enumerate}
\item For each challenge $h_i$ that the cheating verifier $\B$ sends to the honest prover $\P$ in step (1) of the protocol, query the CDH oracle with $h_i$.  Run $\B$ on the set of outputs $\{h_i^x\}$.
\item Simulate the honest verifier $\V$ by querying the challenge oracle $C()$.  Send the output $r$ as input to the cheating prover $\A$.
\item Output $t$, the element of $\G_1$ sent by the cheating prover $\A$ in step (2) of the protocol.
\end{enumerate}
If $(\A,\B)$ successfully breaks the identification scheme, then the element $t$ satisfies $e(g,t) = e(g^a,r)$, and thus by the bilinearity of the pairing, $t = r^a$.  The probability of success of $\C$ is thus at least $\epsilon$.  Furthermore, $\C$ makes at most $q$ queries to the CDH oracle and runs in time $t + O(1)$.
\end{proof}

\section{Identification scheme based on the Strong Diffie-Hellman assumption}
\label{s:sdh}

Protocol \ref{p:cdhid} is very efficient, requiring an exchange of two elements of $\G_1$, one exponentiation for the prover, and two pairing computations for the verifier.  The one-more-CDH assumption required to prove the scheme's security seems reasonable, especially given that similar assumptions are used in the security proofs of two well-known identification schemes \cite{bp}.  However, the fact that the one-more-CDH assumption has not previously appeared in the literature may give one pause, as it is generally not advisable to introduce new assumptions about computational difficulty.  Thus we would like to find an identification scheme that is as efficient as Protocol \ref{p:cdhid} but requires a weaker security assumption, or at least one that is more widely believed to hold for the groups used in implementations.

The difficulty in adapting the BLS signature scheme into an identification scheme resulted from the random oracle nature of the security proof.  Thus we may have more success if we try to adapt a signature scheme that does not require random oracles for its security.  Boneh and Boyen \cite{bb} have devised such a scheme; a full description of the scheme and the theorem describing its security can be found in Appendix \ref{a:signatures}.  The security rests on an assumption known as the {\it Strong Diffie-Hellman assumption}. 

\begin{defn}[{\cite[\S 3.2]{bb}}]
\label{d:sdh}
Let $\G$ be a cyclic group of prime order $p$, and let $g$ be a generator.  The {\it $q$-Strong Diffie-Hellman problem} in $\G$ is defined as follows: given a $(q+1)$-tuple $(g,g^x,g^{(x^2)},\ldots,g^{(x^q)})$ as input, output a pair $(c,g^{1/(x+c)})$, where $c \in \Z_p$.  An algorithm $\A$ has advantage $\epsilon$ in solving the $q$-SDH problem in $\G$ if
$$\Pr \left[ \A(g,g^x,g^{(x^2)},\ldots,g^{(x^q)}) = (c,g^{\frac{1}{x+c}}) \right] \geq \epsilon, $$
where the probability is over the choice of $g \in \G$ and $x \in \Z_p^*$.

We say that the {\it $(t,q,\epsilon)$-Strong Diffie-Hellman assumption} holds in $\G$ if there is no algorithm $\A$ that runs in time $t$ and has advantage $\epsilon$ in solving the $q$-SDH problem in $\G$.
\end{defn}

In our protocol based on the Boneh-Boyen scheme, Victor the Verifier sends a random challenge message to Peggy the Prover, which Peggy then signs with her private key.

\begin{prot}
\label{p:sdhid}
Let $\G_1$, $\G_2$ be cyclic groups of prime order $p$, and let $e \colon \G_1 \times \G_1 \to \G_2$ be a cryptographic pairing.  Let $g$ be a generator of $\G_1$.
\begin{description}
\item[Key generation] Pick random $x,y \leftarrow \Z_p^*$, and compute $u \leftarrow g^x$, $v \leftarrow g^y$, and $z \leftarrow e(g,g)$.  The public key is $(u,v,z)$, and Peggy's secret key is $(x,y)$.
\item[Interactive protocol] \ 
\begin{enumerate}
    \item Victor sends Peggy a random $m \in \Z_p^*$.
    \item Peggy chooses a random $r \in \Z_p^*$, computes $\sigma = g^{1/(x+m+yr)}$, and sends Victor $(\sigma,r)$.
    \item Victor computes $e(\sigma,u \cdot g^m \cdot v^r)$.  If the result is equal to $z$ he outputs $1$ (accept); else he outputs $0$ (reject).
\end{enumerate}
\end{description}
\end{prot}

\begin{thm}
Suppose the $(q',t',\epsilon')$-SDH assumption holds in $\G_1$.  Then Protocol \ref{p:sdhid} defines a $(t,q,\epsilon)$-identification scheme, provided that 
$$ \begin{array}{cccc}
q \leq q', & \epsilon \geq {\displaystyle  2 \epsilon' \cdot \left( \frac{p}{p-q} \right) + \frac{2q}{p-q} }
\approx 2 \epsilon' & \mbox{and} & t \leq t' - \Theta(q'^2T),
\end{array} $$
where $T$ is the maximum time for an exponentiation in $\G_1$.
\end{thm}

\begin{proof}
We first check the viability condition.  If Peggy and Victor both follow the protocol, then Victor will always accept, since
$$e(\sigma,u \cdot g^m \cdot v^r) = e(g^{1/(x+m+yr)},g^x \cdot g^m \cdot g^{yr}) = e(g,g) = z$$
by bilinearity of $e$.
To check the soundness condition, given an attacker $(\A,\B)$ that $(t,q,\epsilon)$-breaks the scheme (in the sense of Definition \ref{d:idscheme}), we can define an attacker $\C$ that $(t + O(1),q,\epsilon')$-breaks the Boneh-Boyen signature scheme, where $\epsilon' = \epsilon(1 - q/p)$.  The reduction is identical to that in the proof of Theorem \ref{t:blsid}, and we choose not to repeat the details.
\end{proof}

\section{Identification scheme based on pairing as a one-way function}
\label{s:owf}

The identification scheme of Protocol \ref{p:sdhid} is less efficient than that of Protocol \ref{p:cdhid}, requiring both more bandwidth and more computation.  However, the assumption required to prove security is weaker for the former, implying a tradeoff between efficiency and security.  One may ask how far we can carry this tradeoff: what is the weakest possible assumption necessary for a secure identification scheme?  We now propose a scheme whose proof of security rests solely on the assumption that the pairing $e \colon \G_1 \times \G_1 \to \G_2$ is a one-way function when one argument is fixed.  This assumption is weaker than both Computational Diffie-Hellman in $\G_1$ and Decision Diffie-Hellman in $\G_2$, both of which are standard assumptions that have been used to prove the security of a wide variety of cryptosystems.

When we say than a pairing is a one-way function, we mean that given $g \in \G_1$ and $y \in \G_2$, it is hard to invert the pairing; that is, to find an element $h \in \G_1$ such that $e(g,h) = y$.

\begin{defn}
\label{d:owp}
Let $e \colon \G_1 \times \G_1 \to \G_2$ be a cryptographic pairing.  We say that $e$ is a {\it $(t,\epsilon)$-one-way pairing} if for any algorithm $\A$ that takes as input $g \in \G_1$ and $x \in \G_2$, produces as output an element of $\G_1$, and runs in time at most $t$,
$$ \Pr \left[e(g, \A(g,x)) = x \right] < \epsilon, $$
where the probability is taken over the possible values of $g$ and $x$.  Given any such $\A$, we say that $\A$ {\it inverts the pairing} with probability at most $\epsilon$.
\end{defn}

To support our claim that one-wayness of pairings is a weak assumption, we note that inverting a pairing is no easier than solving either the Computational Diffie-Hellman problem in $\G_1$ or the Decision Diffie-Hellman problem in $\G_2$.  Indeed, solving the equation $e(g,h) = e(g^a,g^b)$ for $h$ solves the CDH problem for $(g,g^a,g^b)$ in $\G_1$, and solving the equations $e(g,h_i) = z_i$ for $h_i$ given $z_i \in \{z,z^a,z^b,z^c\}$ allows us to use the pairing $e$ to determine whether $z^{ab} = z^c$ in $\G_2$.  For precise statements and proofs of these facts, see Appendix \ref{a:owp}.

Now that we are confident that inverting a pairing is a sufficiently hard problem, we forge onward and define an identification scheme based on the difficulty of inverting a pairing.

\begin{prot}
\label{p:owfid}
Let $\G_1$, $\G_2$ be cyclic groups of prime order $p$, and let $e \colon \G_1 \times \G_1 \to \G_2$ be a cryptographic pairing.
\begin{description}
\item[Key generation] Pick random $P,Q \leftarrow \G_1$, random $y \leftarrow \G_1$, and random $s \leftarrow \Z_p^*$.  Compute $v \leftarrow e(P,Q)^{-1} \cdot y^{-s} \in \G_2$.  The public key is $(P,y,v)$, and Peggy's secret key is $(Q,s)$.
\item[Interactive protocol] \ 
\begin{enumerate}
    \item Peggy chooses random $R \leftarrow \G_1$ and $r \leftarrow \Z_p$, and sends Victor $x = e(P,R) \cdot y^r \in \G_2$.
    \item Victor sends Peggy a random $m \in \Z_p^*$.
    \item Peggy computes $T = R \cdot Q^m \in \G_1$ and $a = r + ms \in \Z_p$, and sends Victor $(T,a)$.
    \item Victor computes $e(P,T) \cdot y^a \cdot v^m \in \G_2$.  If the result is equal to $x$ he outputs $1$ (accept); else he outputs $0$ (reject).
\end{enumerate}
\end{description}
\end{prot}

\begin{rem}
\label{r:viable}
It is easy to see that this protocol is viable: if Peggy and Victor both follow the protocol, Victor will always output $1$, since
\begin{eqnarray*}
e(P,T) \cdot y^a \cdot v^e & = & e(P,R \cdot Q^m) \cdot y^{r+ms} \cdot (e(P,Q)^{-1} \cdot y^{-s})^m \\
& = & e(P,R) \cdot e(P,Q)^m \cdot y^{r+ms} \cdot e(P,Q)^{-m} \cdot y^{-ms} \\
& = & e(P,R) \cdot y^r \\
& = & x.
\end{eqnarray*}
\end{rem}

Showing security is a trickier matter.  Our proof uses the ``heavy row'' technique introduced by Feige, Fiat, and Shamir \cite{ffs} in their seminal paper on proofs of identity.  The proof closely follows those of Okamoto's schemes \cite{ok} based on the discrete logarithm and RSA inversion.  We state the theorem below and give a sketch of the proof; the full proof can be found in Appendix \ref{a:owfproof}.

\begin{thm}
\label{t:owfid}
Suppose $e \colon \G_1 \times \G_1 \to \G_2$ is a $(t',\epsilon')$-one-way pairing, where $\epsilon' > 3/16$ and $p = \abs{\G_1} = \abs{\G_2} \geq 17$.  Then Protocol \ref{p:owfid} is a $(t,q,\epsilon)$-identification scheme, provided that
\begin{eqnarray*}
\epsilon > \frac{2}{p} & \mbox{and} & 
c_0 + \frac{3(t + c_s q)}{\epsilon} \leq t' 
\end{eqnarray*}
for some constants $c_0$, $c_s$ depending on $\G_1$, $\G_2$, and the pairing $e$.
\end{thm}

\begin{proof}[{\bf Proof (sketch)}]
In Remark \ref{r:viable} we demonstrated the viability condition of Definition \ref{d:idscheme}, so we need only show the security condition.  We suppose there is an algorithm $(\A,\B)$ that breaks Protocol \ref{p:owfid}, and construct an algorithm $\C$ that tries to invert the pairing.  Given $P \in \G_1$ and $y \in \G_2$, we simulate Protocol \ref{p:owfid} using $(P,y)$ as the public key and our own randomly chosen private key $(Q^*,s^*)$.  Successful execution of the algorithm $(\A,\B)$ on this instance of the protocol gives a valid interaction between the cheating prover $\A$ and the honest verifier $\V$.  If we run the algorithm again and use the same random coins in the algorithm $(\A,\B)$, the ``heavy row'' lemma tells us that we will, with high probability, find a second valid interaction between $\A$ and $\V$.  From the transcripts of these two interactions we can compute $X \in \G_1$ such that $e(P,X) = y$, and we have inverted the pairing.

The specific description of the algorithm $\C$ is as follows:
\begin{enumerate}
\item Given input $P \in \G_1$ and $y \in \G_2$, choose random $Q^* \in \G_1$ and $s^* \in \Z_p$, and compute $v = e(P,Q^*)^{-1}y^{-s}$.
\item Simulate Protocol \ref{p:owfid} with $(P,y,v)$ as the public key and $(Q^*,s^*)$ as the private key.
\item Run $(\A,\B)$ on the simulated protocol $1/\epsilon$ times.  If the attack succeeds, record $R_{\A\B}$ (the random coins of $(\A,\B)$) and the transcript $(x,m,T,a)$.
\item Run $(\A,\B)$ on the simulated protocol $2/\epsilon$ times, using $R_{\A\B}$ as the random coins.  If the attack succeeds, record the transcript $(x,m',T',a')$.
\item Let $Q = (T/T')^{1/(m-m')} \in \G_1$ and $s = (a-a')/(m-m') \in \Z_p$.  Output 
    $$Z = \left( {Q}/{Q^*}\right)^{1/(s^*-s)}.$$
\end{enumerate}

If steps (3) and (4) succeed and $(Q,s) \neq (Q^*,s^*)$, then step (5) outputs a $Z$ such that $e(P,Z) = y$, and we have inverted the pairing.  Since the probability of success of $(A,B)$ is $\epsilon$, step (3) succeeds with constant probability.  Furthermore, if $\epsilon > 2/p$, then for at least half of the choices of $R_{\A\B}$, the probability of success of $(A,B)$ given the random coins $R_{\A\B}$ is at least $\epsilon/2$.  (This is the ``heavy row'' lemma; see Appendix \ref{a:owfproof} for details.)  Thus step (4) succeeds with constant probability at least half of the time.  Finally, the pairs $(Q,s)$ and $(Q^*,s^*)$ cannot be distinguished even by an infinitely powerful cheating algorithm, so the probability that $(Q,s) \neq (Q^*,s^*)$ is nearly $1$.  When we calculate these probabilities more precisely, we find that the probability of success of $\C$ is at least $3/16$.

Finally, we analyze the running time of $\C$.  If $c_s$ is the time taken to simulate the protocol with the private key $(Q^*,s^*)$, then each iteration of steps (3) and (4) takes time $t + c_s q$, so those two steps take time $3(t+c_s q)/\epsilon$.  Steps (1) and (5) take a constant amount of time, say $c_0$, so the total running time is $c_0 + 3(t+c_s q)/\epsilon$.  
\end{proof}

The assumption $p \geq 17$ in Theorem \ref{t:owfid} is trivial, since in cryptographic applications $p \approx 2^{160}$.  However, the assumption that $e$ is a $(t',\epsilon')$-one-way pairing with $\epsilon' > 3/16$ is a bit stronger than we would like.  If we remove both of these conditions we get the following reduction:

\begin{cor}
\label{c:owfid}
Suppose $e \colon \G_1 \times \G_1 \to \G_2$ is a $(t',\epsilon')$-one-way pairing.  Then Protocol \ref{p:owfid} is a $(t,q,\epsilon)$-identification scheme, provided that
$$\begin{array}{ccc}
{\displaystyle \epsilon \geq \max \left\{ 3 \sqrt{\epsilon'}, \frac{2}{p} \right\}} & \mbox{and} &
{\displaystyle t \leq \frac{t'}{2} - c_0 - c_sq},
\end{array} $$
for some constants $c_0$, $c_s$ depending on $\G_1$, $\G_2$, and the pairing $e$.
\end{cor}

The reduction is the same as in the proof of Theorem \ref{t:owfid}, except we don't iterate steps (3) and (4) of algorithm $\C$.  For full details, see Appendix \ref{a:owfproof}.

\section{Other identification schemes}
\label{s:other}

While there have been several pairing-based identification schemes proposed in the literature, none of these have been given full proofs of security with polynomial-time reductions.  The first such scheme, proposed by Kim and Kim \cite{kk} and based on the Gap Diffie-Hellman problem, was shown to be breakable in constant time by any adversary knowing only the public key.  Yao, Wang, and Wang \cite{yww} proposed a modification of the scheme and proved it to be secure if the Gap Diffie-Hellman problem (cf.\ Remark \ref{r:gdh}) is hard.  However, their reduction requires exponential time, and thus the proof is unsatisfactory.  We will therefore not consider these two schemes when comparing the various pairing-based identification schemes.

More recently, two pairing-based identification schemes have been proposed that appear to be more promising.  Shao, Cao, and Lu \cite{scl} have proposed a scheme very similar to our Protocol \ref{p:sdhid}, based on the Boneh-Boyen signature scheme.  The authors claim that the scheme's security depends on the intractability of the Strong Diffie-Hellman problem, but they do not give a proof, and we have not been able to come up with a reduction.  The scheme is as follows:

\begin{prot}[\cite{scl}]
\label{p:scl}
Let $\G_1$, $\G_2$ be cyclic groups of prime order $p$, and let $e \colon \G_1 \times \G_1 \to \G_2$ be a cryptographic pairing.
\begin{description}
\item[Key generation] Pick random $g \leftarrow \G_1$ and $x \leftarrow \Z_p^*$, and compute $v \leftarrow g^x \in \G_1$ and $z \leftarrow e(g,g) \in \G_2$.  The public key is $(g,v,z)$, and Peggy's secret key is $x$.
\item[Interactive protocol] \ 
\begin{enumerate}
    \item Peggy chooses a random $w \in \Z_p^*$ and sends Victor $\tau = g^w$.
    \item Victor sends Peggy a random $r \in \Z_p^*$.
    \item Peggy sends Victor $\sigma = g^{1/(xr+w)}$.
    \item Victor computes $e(\sigma,\tau \cdot v^r)$.  If the result is equal to $z$ he outputs $1$ (accept); else he outputs $0$ (reject).
\end{enumerate}
\end{description}
\end{prot}

\begin{conj}
Suppose there exists an algorithm $(\A,\B)$ that $(t,q,\epsilon)$-breaks Protocol \ref{p:scl}.  Then there is an algorithm $\C$ that runs in time polynomial in $t$ and $q$ and succeeds in solving the Strong Diffie-Hellman problem with probability polynomial in $\epsilon$.
\end{conj}

The final pairing-based identification scheme we consider was proposed by Hufschmitt, Lefranc, and Sibert \cite{hls}.  The scheme is similar to our Protocol \ref{p:owfid}.  

\begin{prot}[\cite{hls}]
\label{p:hls}
Let $\G_1$, $\G_2$ be cyclic groups of prime order $p$, and let $e \colon \G_1 \times \G_1 \to \G_2$ be a cryptographic pairing.
\begin{description}
\item[Key generation] Pick random $P \leftarrow \G_1$ and $a,b \leftarrow \Z_p^*$, and compute $R \leftarrow P^a, S \leftarrow P^b, Q \leftarrow P^{ab} \in \G_1$ and $z \leftarrow e(P,P), v \leftarrow e(P,P)^{ab} = e(P,Q) \in \G_2$.  The public key is $(P,R,S,v,z)$, and Peggy's secret key is $Q$.
\item[Interactive protocol] \ 
\begin{enumerate}
    \item Peggy sends Victor a random $r \in \Z_p^*$ and sends Victor $w = z^r = e(P,P)^r$.
    \item Victor sends Peggy a random $c \in \Z_p^*$.
    \item Peggy sends Victor $\sigma = P^r \cdot Q^c$.
    \item Victor computes $e(P,\sigma)$ and $w \cdot v^c$ in $\G_2$.  If the two are equal he outputs $1$ (accept); else he outputs $0$ (reject).
\end{enumerate}
\end{description}
\end{prot}

Hufschmitt, Lefranc, and Sibert describe a proof of security of their scheme against a ``passive'' attack involving only a cheating prover $\A$.  They assert that if such an attacker breaks Protocol \ref{p:hls}, then this attacker can be used to solve the Gap Diffie-Hellman problem (cf.\ Remark \ref{r:gdh}), which is (by definition) equivalent to solving the Computational Diffie-Hellman problem in $\G_1$.  

One flaw in the design of Protocol \ref{p:hls} is that the scheme does not make use of the public parameters $R = P^a$ and $S = P^b$, and it appears that they are only included to allow us to reduce breaking the protocol to breaking the Computational Diffie-Hellman problem in $\G_1$.  If we ignore these two parameters, then the passive attacker $\A$ can be used to invert the pairing $e$, and thus the relevant computational assumption is not CDH but the weaker assumption that $e$ is a one-way pairing.

A more serious flaw is that while Protocol \ref{p:hls} appears to be secure against passive attacks, our definition of security (\ref{d:idscheme}) considers an ``active'' attack, which involves a cheating prover $\A$ as well as a cheating verifier $\B$ who tries to gain information by interacting with Peggy, the honest prover.  The protocol's authors do not consider such an attack, and we have not yet found a security assumption under which the scheme is secure.  We conjecture that since the scheme is of the same general format as the Schnorr and Guillou-Quisquater schemes (\cite{sch}, \cite{gq}), the assumption required for security of Protocol \ref{p:hls} will be similar to the assumptions required for the Schnorr and GQ schemes.  The latter are the ``one-more discrete logarithm'' and ``one-more RSA inversion'' assumptions considered by Bellare and Palacio \cite{bp}, so we expect that an analgous ``one-more'' assumption will allow for a proof of security of Protocol \ref{p:hls}.

\section{Comparison of identification schemes}
\label{s:comparison}

We now compare the various identification schemes we have presented in terms of bandwidth and computation required for one iteration of each protocol.  The results are summarized in Table 1. 

\begin{table}[hbt]
\label{t:compare}
\begin{tabular}{|c|c|ccc|ccc|}
\hline
ID & Security & \multicolumn{3}{|c|}{Bandwidth} & \multicolumn{3}{|c|}{Computation} \\
Scheme  & Assumption & $\G_1$ & $\G_2$ & $\Z_p$ & $\G_1$ exp. & $\G_2$ exp. & Pairings \\
\hline
\ref{p:blsid} & CDH in $\G_1$ (ROM)    & 1 & 0 & $1^*$ & 1P & 0 & 2V \\
\ref{p:cdhid} & one-more-CDH      & 2 & 0 & 0 & 1P & 0 & 2V \\
\ref{p:sdhid} & SDH in $\G_1$     & 1 & 0 & 2 & 1P, 2V & 0 & 1V \\
\ref{p:owfid} & $e$ is one-way    & 1 & 1 & 2 & 1P & 1P, 2V & 1P, 1V \\
\ref{p:scl}   & SDH in $\G_1$(?)  & 2 & 0 & 1 & 2P, 1V & 0 & 1V \\
\ref{p:hls}   & ??? & 1 & 1 & 1 & 2P & 1P, 1V & 1V \\
\hline
\end{tabular}
\vspace{8pt}
\caption{Comparison of proposed identification schemes.  The Bandwidth column indicates the number of elements of $\G_1$, $\G_2$, and $\Z_p$ exchanged during one instance of the protocol.  The Computation column indicates how many exponentiations in $\G_1$, exponentiations in $\G_2$, and pairing computations the Prover and Verifier must execute during one instance of the protocol.  We note that the security proof of Protocol \ref{p:blsid} is in the Random Oracle Model.  The entry $1^*$ represents an element of $\bit{n}$; in practice $2^n$ will be around the size of $p$.}
\end{table}

Currently, the only pairings used in cryptographic applications are derived from the Weil and Tate pairings on elliptic curves over finite fields $\F_q$.  These pairings map from the elliptic curve group $E(\F_q)$ to some extension field $\F_{q^k}$; the parameter $k$ is called the {\it embedding degree} of the curve $E$.  For the pairing to be useful, it is necessary that the discrete logarithm problems in $E(\F_q)$ and $\F_{q^k}$ are both hard.  Given current discrete logarithm algorithms, $q \sim 2^{160}$ and $k \sim 2^{1024}$ appear to be reasonable choices for the parameters.

We now assume that $\G_1 = E(\F_q)$, $\G_2 = \F_{q^k}$, and $p \approx q$.  An element $P$ of $E(\F_q)$ can be represented by an element of $\F_q$ corresponding to the $x$-coordinate of $P$, plus one bit for the sign of the $y$-coordinate.  Thus elements of $\G_1$ and $\Z_p$ are of about the same size ($\log_2 p$ bits), while elements of $\G_2$ will be $k$ times as large.  Therefore if minimizing bandwidth is a primary concern, one of Protocols \ref{p:blsid} or \ref{p:cdhid} should be used.  Protocols \ref{p:owfid} and \ref{p:hls} require an element of $\G_2$ to be transmitted, so they should be avoided.

If minimizing computational time is a primary concern, we will wish to minimize pairing computation and perform as few exponentiations as possible in the larger group.  Thus Protocols \ref{p:sdhid} and \ref{p:scl} are ideal for this application.  If we only care about minimizing the Prover's computational time, as in a smart card application, then one of Protocols \ref{p:blsid}, \ref{p:cdhid}, or \ref{p:sdhid} will be best.  However, Protocol \ref{p:blsid} may be less preferable since the prover and verifier must each compute a hash function in addition to performing the group computations.

Finally, if security is the foremost concern, then we should choose a scheme whose proof requires the weakest security assumption.  Table 2 shows the implications between the various computational assumptions used to prove security of our protocols.  We see that the weakest assumption is that the pairing is a one-way function.  Protocol \ref{p:owfid} is based on this assumption, so this scheme is the most secure.  

\begin{table}[htb]
\[ \xymatrix{
& \txt{$e \colon \G_1 \times \G_1 \to \G_2$ \\ is a one-way pairing \\ (Definition \ref{d:owp})} \ar@{<=}[d] &  \\
& \txt{CDH in $\G_1$ \\ (Definition \ref{d:cdh})} \ar@{<=}[dl] \ar@{<=}[dr] & \\
 \txt{SDH in $\G_1$ \\ (Definition \ref{d:sdh})} & &  \txt{one-more-CDH in $\G_1$ \\ (Definition \ref{d:omcdh})}
} \]
\caption[Figure 1]{Implications between various computational assumptions.}
\end{table}

\section{Conclusion}

We have presented four new identification schemes based on pairings, and proved their security given various computational assumptions.  Each of our schemes is at least as efficient and/or secure as any scheme currently in the literature.  Our main contribution is Protocol \ref{p:owfid}, a scheme which is secure if the pairing in question is a one-way function; this assumption is weaker than that made for any other pairing-based scheme currently in the literature.

For another of our schemes, Protocol \ref{p:cdhid}, we introduced an assumption called the ``one-more-CDH'' assumption, analogous to the ``one-more-discrete-log'' and ``one-more-RSA-inversion'' assumptions, and proved our scheme secure under this assumption.  An important open question is what relation this assumption has to other computational assumptions in the literature.

\appendix

\section{Pairing-based signature schemes}
\label{a:signatures}

In this appendix, we describe the pairing-based signature schemes that are the basis for the identification schemes defined in Protocols \ref{p:blsid} and \ref{p:sdhid}.  We give a definition of security for signature schemes and state the security theorems for the two protocols in question.

We first describe the pairing-based short signature scheme devised by Boneh, Lynn, and Shacham \cite{bls}, on which our Protocol \ref{p:blsid} is based.  We describe the scheme in terms of a pairing, but the scheme is in fact valid in any group in which the Decision Diffie-Hellman problem is easy and the Computational Diffie-Hellman problem is hard; such a group is called a {\it Gap Diffie-Hellman group}.

\begin{prot}[\cite{bls}]
\label{p:bls}
Let $\G_1$, $\G_2$ be cyclic groups of prime order $p$, and let $e \colon \G_1 \times \G_1 \to \G_2$ be a cryptographic pairing.  Let $g$ be a generator of $\G_1$.  Let $H : \bit{*} \to \G_1$ be a full-domain hash function.
\begin{description}
\item[Key generation] Pick random $x \leftarrow \Z_p$, and compute $v \leftarrow g^x$.  The public key is $v$, and the secret key is $x$.
\item[Signing] Given a secret key $x \in \Z_p$ and a message $M \in \bit{*}$, compute $h \leftarrow H(M)$ and $\sigma \leftarrow h^x$.  The signature is $\sigma \in \G$.
\item[Verification] Given a public key $v \in \G$, a message $M \in \bit{*}$, and a signature $\sigma \in \G$, compute $e(g,\sigma)$ and $e(v,h)$.  If the two are equal, output {\tt valid}; if not, output {\tt invalid}.
\end{description}
\end{prot}

Boneh, Lynn, and Shacham prove the security of their scheme using the following game between a challenger and an adversary $\A$.

\begin{description}
\item[Setup] The challenger runs algorithm $KeyGen$ to optain a public key $PK$ and a private key $SK$.  The adversary $\A$ is given $PK$.
\item[Queries] Proceeding adaptively, $\A$ requests signatures with $PK$ on at most $q_S$ messages of his choice, $M_1,\ldots,M_{q_s} \in \bit{*}$.  The challenger responds to each query with a signature $\sigma_i = Sign(SK,M_i)$.
\item[Output] Eventually, $\A$ outputs a pair $(M,\sigma)$ and wins the game if (1) $M$ is not any of $M_1,\ldots,M_{q_S}$, and (2) $Verify(PK,M,\sigma) = $ {\tt valid}.
\end{description}

The advantage of $\A$, denoted $\Adv(\A)$, is the probability that $\A$ wins the above game, taken over the coin tosses of $KeyGen$ and of $\A$ itself.  We are now ready to define the security of a signature scheme.

\begin{defn}[{\cite[Definition 3.1]{bls}}]
A forger $\A$ {\it $(t,q_S,q_H,\epsilon)$-breaks} a signature scheme if $\A$ runs in time at most $t$, makes at most $q_S$ signature queries and at most $q_H$ queries to a hash function, and $\Adv(\A) > \epsilon$.  A signature scheme is {\it $(t,q_S,q_H,\epsilon)$-existentially unforgeable under adaptive chosen-message attack} if no forger $(t,q_S,q_H,\epsilon)$-breaks it.
\end{defn}

The security of the BLS signature scheme is based on the Computational Diffie-Hellman assumption in the group $\G_1$ (Defintion \ref{d:cdh}).

\begin{thm}[{\cite[Theorem 3.2]{bls}}]
\label{t:bls}
Suppose the $(t',\epsilon')$-Computational Diffie-Hellman assumption holds in $\G_1$.  Then the signature scheme defined in Protocol \ref{p:bls} is $(t,q_S,q_H,\epsilon)$-secure against existential forgery under an adaptive chosen-message attack (in the random oracle model) for all $t$ and $\epsilon$ satisfying
$$\begin{array}{ccc}
\epsilon \geq e(q_S+1)\cdot \epsilon' & \mbox{and} & t \leq t' - c(q_H + 2q_S),
\end{array}$$
where $c$ is a constant that depends on $\G_1$, and $e$ is the base of the natural logarithm.
\end{thm} 

The second signature scheme we describe was devised by Boneh and Boyen \cite{bb}; our identification scheme \ref{p:sdhid} is based on this scheme.  

\begin{prot}[\cite{bb}]
\label{p:bb}
Let $\G_1$, $\G_2$ be cyclic groups of prime order $p$, and let $e \colon \G_1 \times \G_1 \to \G_2$ be a cryptographic pairing.  Let $g$ be a generator of $\G_1$.
\begin{description}
\item[Key generation] Pick random $x,y \leftarrow \Z_p^*$, and compute $u \leftarrow g^x$, $v \leftarrow g^y$, and $z \leftarrow e(g,g)$.  The public key is $(u,v,z)$, and the secret key is $(x,y)$.
\item[Signing] Given a secret key $(x,y) \in (\Z_p^*)^2$, and a message $m \in \Z_p^*$, pick a random $r \in \Z_p^*$ and compute $\sigma \leftarrow g^{1/(x+m+yr)} \in \G_1$, where $1/(x+m+yr)$ is computed modulo $p$.  In the (unlikely) event that $x + m + yr = 0 \pmod{p}$, try again with a different random $r$.  The signature is $(\sigma,r)$ .
\item[Verification] Given a public key $(u,v,z) \in \G_1^2 \times \G_2$, a message $m \in \Z_p^*$, and a signature $(\sigma,r) \in \G_1 \times \Z_p^*$, compute $e(\sigma,u\cdot g^m \cdot v^r)$.  If the result is equal to $z$ output {\tt valid}; if not, output {\tt invalid}.
\end{description}
\end{prot}

The security of the Boneh-Boyen scheme is based on the Strong Diffie-Hellman assumption (Definition \ref{d:sdh}).  The relevant fact about the proof of security is that it gives a tight reduction without using the random oracle model.

\begin{thm}[{\cite[Theorem 3.1]{bb}}]
Suppose the $(q,t',\epsilon')$-SDH assumption holds in $\G_1$.  Then the signature scheme defined by Protocol \ref{p:bb} is $(t,q_s,\epsilon)$-secure against existential forgery under adaptive chosen message attack, provided that 
$$ \begin{array}{cccc}
q_s \leq q, & \epsilon \leq 2 \left(\epsilon' + q_S/p \right) \approx 2 \epsilon' & \mbox{and} & t \leq t' - \Theta(q^2T),
\end{array} $$
where $T$ is the maximum time for an exponentiation in $\G_1$.
\end{thm}

\section{Security of Protocol \ref{p:blsid}}
\label{a:blsproof}

\begin{proof}[{\bf Proof of Theorem \ref{t:blsid}}]
If Peggy and Victor follow the protocol, then Protocol \ref{p:blsid} satisfies the viability condition of Definition \ref{d:idscheme}, since
$$e(g,\sigma) = e(g,h^x) = e(g,h)^x = e(g^x,h) = e(v,h)$$
by bilinearity of $e$.

To show the security condition, it suffices to show that if the BLS signature scheme (Protocol \ref{p:bls}) is $(t',q,r,\epsilon')$-secure against existential forgery under an adaptive chosen-message attack, then Protocol \ref{p:blsid} is a $(t,q,r,\epsilon)$ identification scheme, provided that
$$ \begin{array}{ccc}
\epsilon \geq {\displaystyle \left( \frac{2^n}{2^n-q} \right) \cdot \epsilon'}  & \mbox{and} & t \leq t' - c
\end{array}$$
for some constant $c$ depending on the groups and pairing used.  If we give a reduction from the identification scheme to the signature scheme with these bounds, then the security theorem for the BLS signature scheme (Theorem \ref{t:bls}) implies that there is a reduction from the identification scheme to the CDH problem in $\G_1$ with the stated bounds.

To construct the specified reduction, we now suppose that $(\A,\B)$ is a pair of algorithms that $(t,q,r,\epsilon)$-breaks the scheme (in the sense of Definition \ref{d:idscheme}) for a given public/private-key pair.  Define an attacker $\C$ on the BLS scheme with the same public and private keys, as follows:  
\begin{enumerate}
\item For each $M_i$ that the cheating verifier $\B$ sends to the honest prover $\P$, have $\C$ request a signature on $M_i$.  Run $\B$ on the output.
\item Simulate the honest verifier $\V$ by choosing a random $M$ and sending $M$ as input to the cheating prover $\A$.
\item Output the pair $(M,\tau)$, where $\tau \in \G_1$ is the element that the cheating prover $\A$ sends to $\V$.
\end{enumerate}
If $(\A,\V)$ outputs $1$, then the output of algorithm $\C$ is a valid BLS message-signature pair.  Thus if $M$ is distinct from all of the queries $M_i$, then $(M,\tau)$ is a valid forgery.  Since the probability of $(\A,\B)$ simulating the prover $\P$ is at least $\epsilon$ and the probability that $M$ is equal to one of the $M_i$ is $q/2^n$, the probability of forging a signature is at least $(1 - q/2^n) \cdot \epsilon$.  We thus have broken the BLS scheme with an attacker that runs in time $t+c$ for some constant $c$.  The attacker makes $q$ signature queries and $h$ hash queries.
\end{proof}

\section{Hardness of inverting a one-way pairing}
\label{a:owp}

In Section \ref{s:owf} we stated that the assumption that $e \colon \G_1 \times \G_1 \to \G_2$ is a one-way pairing is weaker than both the Computational Diffie-Hellman assumption in $\G_1$ and the Decision Diffie-Hellman assumption in $\G_2$.  We now give precise statements and proofs of these facts.

\begin{prop}
Let $e \colon \G_1 \times \G_1 \to \G_2$ be a cryptographic pairing between groups of order $p$.  Suppose the $(t,\epsilon)$ Computational Diffie-Hellman assumption holds in $\G_1$.  Then $e$ is a $(t - O(1), \epsilon)$-one-way pairing.
\end{prop}

\begin{proof}
Let $\A(g,x)$ be an algorithm that runs in time $t$ and inverts the pairing with probability at least $\epsilon$.  Given a triple $(h,h^a,h^b)$ of elements in $\G_1$, let $y = e(h^a,h^b)$, and run $\A(h,y)$.  Then $\A$ outputs $h^{ab}$ with probability at least $\epsilon$.
\end{proof}

\begin{prop}
Let $e \colon \G_1 \times \G_1 \to \G_2$ be a cryptographic pairing between groups of order $p$.  Suppose the $(t,\epsilon)$-Decision Diffie-Hellman assumption holds in $\G_2$.  Then $e$ is a $(t/\epsilon - O(1), \sqrt[4]{\epsilon})$-one-way pairing.
\end{prop}

\begin{proof}
Let $\A(g,x)$ be an algorithm that runs in time $t$ and inverts the pairing with probability at least $\epsilon$.
We are given a quadruple $\{y,y^a,y^b,y^c\}$ of elements of $\G_2$ and asked to determine if $c = ab \pmod{p}$.  Define algorithm $\B$ as follows.
\begin{enumerate}
\item Choose a random $g \in \G_1$, and compute 
\begin{equation*} \begin{array}{cc}
    h_1  =  \A(g,y), &
    h_2  =  \A(g,y^a), \\
    h_3  =  \A(g,y^b), &
    h_4  =  \A(g,y^c).
\end{array} \end{equation*}
\item Compute $e(h_1,h_4)$ and $e(h_2,h_3)$.  If the two are equal output $1$; else output $0$.
\end{enumerate}
Suppose all four outputs of algorithm $\A$ are correct.  Then $h_2 = h_1^a$, $h_3 = h_1^b$, and $h_4 = h_1^c$.  We therefore have $e(h_1,h_4) = e(h_1,h_1)^c$ and $e(h_2,h_3) = e(h_1,h_1)^{ab}$.  The two are equal if and only if $c = ab \pmod{p}$.  Thus if all four outputs are correct $\B$ gives a correct output to the Decision Diffie-Hellman problem.  The probability that all four outputs are correct is at least $\epsilon^4$, which gives the stated security bound.  Furthermore, $\B$ runs in time $4t + O(1)$.
\end{proof}

\begin{rem}
We can increase the probability of success of $\B$ by iterating the algorithm.  Performing each computation of $h_i$ $\epsilon^{-4}$ times increases the probability of success to a constant; fewer repetitions lead to different time/success ratios.
\end{rem}

\section{Security of Protocol \ref{p:owfid}}
\label{a:owfproof}

In this appendix, we show that Protocol \ref{p:owfid} is secure if we assume that $e$ is a one-way pairing.  The proof adapts Okamoto's arguments for proving security of his two identification schemes \cite{ok}.  We begin the detailed proof by defining a ``heavy row'' and proving some useful lemmas.

\begin{defn}
Let $(\A,\B)$ be an algorithm attacking Protocol \ref{p:owfid}.  Let $R_{\A\B}$ denote the random coins consumed by $(\A,\B)$.  Let $M$ be a matrix summarizing all of the possible outcomes of the cheating prover $\A$ interacting with an honest verifier $\V$, as follows: the rows of $M$ are indexed by the possible choices of $R_{\A\B}$,  the columns of $M$ are indexed by all the possible choices $e$ of the verifier $\V$ in step (2), and the entries are $1$ if $\V$ accepts $\A$'s proof, and $0$ otherwise.

Suppose the probability of success of $(\A,\B)$ (i.e.\ the fraction of $1$'s in $M$) is $\epsilon$.  A row of $M$ is a {\it heavy row} if its fraction of $1$'s is at least $\epsilon/2$.
\end{defn}

\begin{lemma}
\label{l:heavy}
Suppose the success probability of $(\A,\B)$ in attacking Protocol \ref{p:owfid} is at least $2/p$.  Then at least half of the $1$'s in $M$ are located in heavy rows.
\end{lemma}

\begin{proof}
Assume the contrary, i.e.\ at least half the $1$'s in $M$ are located in non-heavy rows.  Then the fraction of $1$'s in all of the non-heavy rows combined is at least $1/p$.  On the other hand, in each non-heavy row the fraction of $1$'s is by definition less than $1/p$, a contradiction.
\end{proof}

\begin{lemma}
\label{l:iterate}
Let $(\A,\B)$ be an algorithm attacking Protocol \ref{p:owfid} that runs in time $t$ and has success probability $\epsilon > 2/p$.  Then there is a algorithm that runs in expected time $O(t/\epsilon)$ and, with probability at least $\frac{1}{2}(1 - \frac{1}{e})^2$ outputs the history of two accepted interactions $(x,m,T,a)$ and $(x,m',T',a')$ of the cheating prover $\A$ with an honest verifier $\V$, where $m \neq m'$.
\end{lemma}

\begin{proof}
We adopt the following two-step ``probing strategy'' (cf.\ \cite{oo}, \cite{ok}) to find two $1$'s in the same row of $M$.
\begin{description}
\item[Step 1] Probe random entries in $M$ to find an entry $a_0$ that is a $1$.  Denote the row in which $a_0$ is located by $M_0$.
\item[Step 2] Probe random entries along $M_0$ to find another entry $a_1$ with $1$.
\end{description}
Let $p_1$ be the success probability of Step 1 after probing $1/\epsilon$ random entries of $M$.  Since the fraction of $1$'s in $M$ is $\epsilon$, we have
$$p_1 \geq 1 - (1 - \epsilon)^{1/\epsilon} > 1 - \frac{1}{e}.$$
Let $p_2$ be the success probability of Step 2 after probing $2/\epsilon$ random entries of $M_0$.  If $M_0$ is a heavy row, then the fraction of $1$'s in $M_0$ is at least $\epsilon/2$, and thus the probability of success is at least
\[ 1 - \left(1 - \frac{\epsilon}{2}\right)^{2/\epsilon} > 1 - \frac{1}{e}. \]
By Lemma \ref{l:heavy}, the probability that $M_0$ is a heavy row is at least $1/2$, and thus $p_2 > \frac{1}{2}(1- \frac{1}{e})$.  Therefore the overall success probability of our strategy is at least $\frac{1}{2}(1- \frac{1}{e})^2$, and the total running time is approximately $3t/\epsilon$.  

If the strategy finds two entries $a_0,a_1$ in the same row of $M$, we output the transcripts $(x,e,T,a)$ and $(x,e',T',a')$ of the interaction between $\A$ and $\V$ when given the random coins corresponding to $a_0$ and $a_1$ respectively.  Since the entries are in the same row, the random coins of $(\A,\B)$ are the same for the two interactions, and thus the first output $x$ is the same for the two interactions.  Since the entries are in different columns, the random coins of $\V$ are different for the two interactions, and thus $m \neq m'$.
\end{proof}

With this setup, we may now prove the security of our identification scheme.

\begin{proof}[{\bf Proof of Theorem \ref{t:owfid}}]
In Remark \ref{r:viable} we demonstrated the viability condition of Definition \ref{d:idscheme}, so we need only show the security condition.  Suppose $(\A,\B)$ is an algorithm that runs in time $t$ and attacks Protocol \ref{p:owfid} with success probability $\epsilon > 2/p$.  Define an algorithm $\C$ that attempts to invert the pairing, as follows:
\begin{enumerate}
\item Given input $P \in \G_1$ and $y \in \G_2$, choose random $Q^* \in \G_1$ and $s^* \in \Z_p$, and compute $v = e(P,Q^*)^{-1}y^{-s}$.
\item Simulate Protocol \ref{p:owfid} with $(P,y,v)$ as the public key and $(Q^*,s^*)$ as the private key.
\item Run $(\A,\B)$ on the simulated protocol $1/\epsilon$ times.  If the attack succeeds, record $R_{\A\B}$ (the random coins of $(\A,\B)$) and the transcript $(x,m,T,a)$.
\item Run $(\A,\B)$ on the simulated protocol $2/\epsilon$ times, using $R_{\A\B}$ as the random coins.  If the attack succeeds, record the transcript $(x,m',T',a')$.
\item Let $Q = (T/T')^{1/(m-m')} \in \G_1$ and $s = (a-a')/(m-m') \in \Z_p$.  Output 
    $$Z = \left( {Q}/{Q^*}\right)^{1/(s^*-s)}.$$
\end{enumerate}

We now analyze the algorithm $\C$.  By Lemma \ref{l:iterate}, the probability that steps (3) and (4) both succeed and output valid transcripts with $m \neq m'$ is at least $\frac{1}{2}(1- \frac{1}{e})^2$.  
We now claim that if steps (3) and (4) both succeed, then $(Q,s) \neq (Q^*,s^*)$ with probability almost $1$.  To prove this, we show that if $(Q,s)$ and $(Q^*,s^*)$ are both valid private keys for the public key $(P,y,v)$, then even an infinitely powerful cheater $\B$ cannot distinguish the two solely from his interaction with an honest prover $\P$.  The condition $(Q,s)$ and $(Q^*,s^*)$ both being valid private keys for the public key $(P,y,v)$ implies that
\begin{equation}
\label{eq:keys}
e(P,Q) \cdot y^s = e(P,Q^*) \cdot y^{s^*}.
\end{equation}
Let $R^* = R + (Q - Q^*)^m \in \G_1$ and $r^* = r + m(s - s^*) \in \Z_p$.  Then the following relations hold:
\begin{equation*}
\begin{split}
e(P,R)\cdot y^r  &=  x = e(P,R^*) \cdot y^{r^*} \\
R + Q^m  &=  T =  R^* + Q^{*m} \\
r + ms  &=  a = r^* + ms^*
\end{split}
\end{equation*}
Furthermore, for given $(Q,Q^*,s,s^*,m)$, the distribution of $(R,r)$ is identical to that of $(R^*,r^*)$.  Since the cheating verifier $\B$ receives only $(x,T,a)$ from the honest prover $\P$, we see that there is no way for $\B$ to determine which private key was used.  Since there are $p$ possible pairs $(Q,s)$ satisfying $e(P,Q)^{-1}y^{-s} = v$, the probability that $(Q,s) \neq (Q^*,s^*)$ is $(p-1)/p$, or nearly $1$.

We now show that if steps (3) and (4) succeed and $(Q,s) \neq (Q^*,s^*)$, then step (5) outputs a $Z$ such that $e(P,Z) = y$.  We first note that if $(Q,s) \neq (Q^*,s^*)$, then equation (\ref{eq:keys}) implies that $Q \neq Q^*$ and $s \neq s^*$, so $Z$ is well-defined.  Since $x$ is the same in both transcripts, we have
$$e(P,T)\cdot y^a\cdot v^m = e(P,T') \cdot y^{a'} \cdot v^{m'}.$$
By the bilinearity of the pairing, this implies that
$$e(P,T/T') \cdot y^{a-a'} = v^{m'-m},$$
so by definition of $Q$ and $s$ we have
$$e(P,Q^{m-m'}) \cdot y^{s(m-m')} = v^{m'-m} $$
Raising the whole equation to the power $1/(m-m')$ and applying the definition $v = e(P,Q^*)^{-1} \cdot y^{-s}$ gives
$$e(P,Q) \cdot y^s = e(P,Q^*) y^{s^*}. $$
Again using the bilinearity of the pairing, this gives us
$$e(P,Q/Q^*) = y^{s^*-s}, $$
and raising both sides to the power $1/({s^*}-s)$ gives
$$e(P,Z) = y,$$
as desired.

Finally, we analyze the running time and success probability of $\C$.  If $c_s$ is the time taken to simulate the protocol with the private key $(Q^*,s^*)$, then each iteration of steps (3) and (4) takes time $t + c_s q$, so those two steps take time $3(t+c_s q)/\epsilon$.  Steps (1) and (5) take a constant amount of time, say $c_0$, so the total running time is $c_0 + 3(t+c_s q)/\epsilon$.  By Lemma \ref{l:iterate} and our computations above, if steps (3) and (4) succeed and $(Q,s) \neq (Q^*,s^*)$, then step (5) outputs a valid $Z$.  The probability of the former is at least $\frac{1}{2}(1- \frac{1}{e})$, while the probability of the latter is $(p-1)/p$.  If $p \geq 17$ then the simultaneous probability of the two events is at least $3/16$.  Thus our reduction gives the stated bounds.
\end{proof}

Finally, we give the detailed proof of Corollary \ref{c:owfid}, a security theorem for Protocol \ref{p:owfid} that does not require any assumptions on the security parameter $\epsilon'$ for the one-way pairing or the size of $p$, the order of $\G_1$ and $\G_2$.

\begin{proof}[{\bf Proof of Corollary \ref{c:owfid}}]
The reduction is the same as in the proof of Theorem \ref{t:owfid}, except we don't iterate steps (3) and (4) of algorithm $\C$.  Then the success probability of step (3) is $\epsilon$.  By Lemma \ref{l:heavy} the entry of the summary matrix $M$ corresponding to the output of step (3) is in a heavy row with probability at least $1/2$, and if this is the case then the success probability of step (4) is at least $\epsilon/2$.  The success probability of step (5) is still $(p-1)/p$, which is at least $1/2$ since $p \geq 2$.  Thus the total success probability $\pi$ of the algorithm satisfies
$$\pi \geq \epsilon \cdot  \frac{1}{2} \cdot \frac{\epsilon}{2} \cdot \frac{1}{2} > \frac{\epsilon^2}{9}.$$  
The algorithm takes time $2(t + c_s q) + 2c_0$, where $c_s$ is the time taken to simulate the protocol and $2c_0$ is the time taken to perform the computations in steps (1) and (5).  Thus our reduction gives the stated bounds.
\end{proof}

\end{document}